\documentclass{article}

\usepackage{cite}
\usepackage{amsmath,amssymb,amsfonts}
\usepackage{algorithm}
\usepackage{algorithmic}
\usepackage{graphicx}
\usepackage{textcomp}
\usepackage{xcolor}
\usepackage{url}
\usepackage{authblk}

\begin{document}

\title{From Idle to Urgent: A Resource-Harvested HPC Workflow for High-Fidelity Seismic Estimation}

\author[1]{Tsuyoshi Ichimura}
\author[1,2]{Kohei Fujita}
\author[1]{Hideaki Ito} 
\author[1]{Wataru Sakurai}
\author[3]{Muneo Hori}
\author[1]{Lalith Maddegedara}
\affil[1]{Earthquake Research Institute and Department of Civil Engineering, The University of Tokyo, Japan}
\affil[2]{RIKEN Center for Computational Science, Japan}
\affil[3]{Japan Agency for Marine-Earth Science and Technology, Japan}

\date{}
\maketitle

\begin{abstract}
We propose an Urgent Interactive HPC workflow that dynamically integrates high-fidelity 3D nonlinear analysis with surrogate neural networks (NNs) to enable rapid decision-making during large-scale earthquakes.
This approach achieves both "Resource Harvesting", which utilizes idle computing capacity during non-emergency periods, and immediate response during crises.
By developing two specialized HPC kernels, the proposed method reduces energy-to-solution by 76\% and improves throughput by 3.7-fold during normal operations to efficiently construct training datasets, while during emergencies, it couples NN-based inverse analysis with physics-based simulations and dynamic refinement to reduce conventional computational costs by over 97.9\%, enabling the generation of highly reliable spatial time-history ground motion distributions within 30 minutes post-earthquake.
\end{abstract}

\section{Introduction}
\label{sct1}

Due to natural forces (e.g., river meandering) and artificial modifications (e.g., cutting and filling), complex shallow 3D ground structures with depths on the order of $10^0$--$10^1$ m are often formed near the surface. 
During large-scale earthquakes, such ground heterogeneity locally amplifies seismic waves, resulting in significantly severe structural damage at specific sites compared to surrounding areas (e.g., \cite{review, pipeline}).
While numerous methods targeting immediate post-earthquake emergency response have been proposed (e.g., \cite{shakemap, caltech, jrisq}), further enhancing the reliability of rapid post-disaster decision-making in dense modern urban areas requires moving beyond coarse ground motion estimates. It is crucial to advance toward high-fidelity, high-resolution spatial time-history ground motion simulations that physically account for localized 3D subsurface structural characteristics.

For instance, considering the Tokyo metropolitan area in Japan (approximately $10^5 \times 10^5\,\mathrm{m}$), there are $10^2$--$10^3$ scattered locations of concern, with approximately size of $10^3 \times 10^3\,\mathrm{m}$ each, featuring complex local ground structures that require high-fidelity evaluations.
Upon the occurrence of a large-scale earthquake, $10^1$--$10^2$ specific sites experiencing strong shaking must be automatically screened from these numerous candidate locations, and detailed spatial ground motion estimations must be completed within a stringent 30-minute urgent decision-making time frame post-event.

Recent advancements in dynamic coupling between observation systems and supercomputers via high-speed networks \cite{super2020}, exemplified by "Edge-to-HPC" and "Superfacility" concepts, are making it technically feasible to collect real-time ground motion observations from affected areas immediately after an earthquake and stage them directly into HPC environments.
However, these collected data consist of sparse observation points on the ground surface, making it difficult to estimate detailed time-history ground motion distributions across the entire region from these data alone.
Estimating ground motion distributions from sparse observations constitutes a challenging inverse problem that generally requires nonlinear optimization---repeatedly executing forward 3D nonlinear ground amplification analyses to search for input bedrock motions consistent with observed waveforms.
This implies performing computationally expensive 3D nonlinear finite-element dynamic analyses at least $10^{2-3}$ times per site for numerous target locations immediately after an earthquake, which is prohibitive due to extreme computational costs within severe time constraints, even when employing state-of-the-art flagship HPC resources.
On the other hand, while pure end-to-end surrogate neural networks (NNs) that infer damage distributions directly from observed waveforms offer fast execution, they carry risks of physical inconsistency and reduced explainability due to black-box properties, leaving critical reliability concerns in life-safety emergency decision-making.

Meanwhile, research on leveraging HPC to overcome computational resource constraints in emergencies and establishing real-time urgent forecasting workflows have been conducted actively in recent years (e.g., \cite{tsunami2020, SC25GB}).
Building upon these prior studies, designing and constructing a computational workflow tailored to the target hardware and system-operational characteristics is expected to drastically reduce the large analytical costs, improve time-to-insight, and enable rapid decision-making support.
To address these challenges, we enable both high energy efficiency during normal operations ("Resource Harvesting") together with immediate responsiveness during crises ("Urgent Computing"), by proposing a computational workflow based on high-fidelity simulations satisfying physics-based equations and integrating rapid surrogate-NN-based inverse estimation and dynamic refinement for drastic reduction in the massive analytical cost.
The remainder of this paper is organized as follows. 
Section~\ref{sct2} presents the overall architecture of the workflow, which combines normal-operation resource harvesting with emergency response and dynamic refinement.
Section~\ref{sct3} details the design and performance evaluation of two specialized core kernels developed to mitigate the computational cost of 3D ground amplification analysis, which represents the main bottleneck of the workflow. 
Section~\ref{sct4} demonstrates the estimation accuracy of the proposed workflow for realistic earthquake scenarios and validates its practical feasibility within an emergency response timeline.

\section{Urgent and Interactive Workflow for Quick Earthquake Response}
\label{sct2}

We present an overview of an urgent and interactive computational workflow designed to estimate high-fidelity spatial time-history ground motion distributions within a realistic time frame for immediate post-event damage estimation and decision support. 
The characteristics of this workflow lie in coupling surrogate-NN-based inverse analysis with accuracy-guaranteed, high-cost physics simulations, by selectively utilizing two specialized core kernels (\texttt{CPUGPU-Harvest} and \texttt{EBEGPU-Urgent}), each optimized at the computer architecture level to align with the distinct resource utilization profiles and power constraints during supercomputer operations under normal and emergency response conditions.

In conventional frameworks without surrogate NNs, inversely estimating input motions consistent with observed data required iterating computationally heavy forward 3D nonlinear ground amplification analyses $10^{2-3}$ times for nonlinear optimization within the tight post-event time frame.
Such an approach is virtually impossible within practical execution times for real-time emergency forecasting.
In contrast, our workflow couples two specialized computational kernels---optimized through high-performance computing techniques---with inverse analysis based on pretrained surrogate NNs.
This dramatically reduces the number of costly 3D analyses conducted during the emergency response phase (down to a minimum of a single run), enabling execution within emergency response time frames. 
Furthermore, by incorporating an on-the-fly dynamic refinement process on residual errors, the workflow delivers rapid, highly reliable insights for disaster prevention and mitigation decision-making.

Here we note that direct inference using pure end-to-end surrogate NNs suffers from inherent drawbacks, including potential physical inconsistencies and limited explainability caused by black-box models.
In our workflow, these limitations are overcome by feeding the inference results into a physics-based simulation at least once---thereby yielding high-fidelity spatial time-history distributions that strictly and consistently reflect strong nonlinear behaviors, such as complex subsurface structural effects, energy dissipation, and wave trapping---and by embedding a dynamic refinement process. Consequently, this workflow ensures the high level of reliability essential for life-safety emergency decision-making.
The details of each phase are described below.

\subsection{Pre-Earthquake Phase: Resource Harvesting}

To generate training datasets for NNs, a vast number of nonlinear ground amplification analyses must be precomputed across numerous target sites.
Since this process requires large energy usage, designing for superior energy-to-solution while accommodating fine-grained, opportunistic scheduling that leverages backfill slots in batch schedulers is important.
In our workflow, we employ a highly efficient core kernel based on heterogeneous computing (\texttt{CPUGPU-Harvest}), designed to effectively utilize both the compute and memory resources of CPUs and GPUs.
This enables automatic and efficient harvesting of idle resources during normal operations across scales ranging from small fragments on a few nodes to large node allocations, thereby establishing an efficient, automated data generation scheme.
Here we note that while training datasets must be regenerated in cases such as when the ground structure information is updated, standard job schedulers can be used to allocate idle resources as each of the training data generation jobs are in sizes of several compute nodes for a short elapsed limit time.

Using the datasets pre-generated through this process, surrogate NNs that inversely infer input bedrock motions from surface ground motion observations are constructed for each site prior to an event (along with hyperparameter optimization).
To enable rapid, on-the-fly fine-tuning (interactive refinement) during the subsequent emergency phase, we adopt a lightweight NN architecture with a constrained parameter count capable of adapting efficiently within a small number of steps (see Appendix for details on the NN architecture).

\subsection{Post-Earthquake Phase: Urgent Response and Interactive Refinement}

Immediately following an earthquake, observed ground motion data are received from surface monitoring stations, and the following two-stage process is dynamically executed for sites where strong shaking was recorded.

\textbf{Stage 1: Rapid 1st-Order Estimation.}
Input bedrock motions are instantaneously inferred using the pretrained NNs. 
Taking these estimated bedrock motions as input, the multi-node GPU parallel kernel (\texttt{EBEGPU-Urgent})---specifically designed to prioritize the minimization of time-to-solution during emergencies---is immediately executed. 
By running the high-fidelity 3D nonlinear ground amplification forward analysis only once, high-fidelity spatial ground motion distributions physically reflecting complex subsurface structures and strong soil nonlinearities are consistently estimated and delivered to decision-makers within the stringent 30-minute decision-making time frame.

\textbf{Stage 2: Interactive Dynamic Refinement.}
The observed surface ground motions are compared with the reproduced waveforms obtained from Stage 1. 
Based on user interface interventions by decision-makers (e.g., requesting priority of analysis for specific targeted areas) or automated thresholding of waveform discrepancies (e.g., relative L1 error exceeding a designated threshold), the results of this 1st-order estimation (pairs of input bedrock motions and surface responses) are immediately fed back into the NN training loop as training data to perform dynamic on-the-fly fine-tuning (an additional training step that incorporates observation-backed data while retaining the search capability derived from random wave training sets; see Appendix for details).
Bedrock input motions are then re-estimated using the updated NN, and the ground amplification analysis is re-executed. 
Through this "interactive adaptive control loop," predictive accuracy is progressively refined until it aligns with observational data, continuously providing updated estimates to decision-makers.

\section{Design and Optimization of HPC Kernels}
\label{sct3}
As the core of the proposed workflow, we present and evaluate the performance of two specialized core kernels: \texttt{CPUGPU-Harvest}, which minimizes energy-to-solution and maximizes throughput during normal operations, and \texttt{EBEGPU-Urgent}, which minimizes latency and time-to-solution during emergencies.
Achieving the detailed time-history ground motion distributions targeted by this workflow requires finite-element simulations using unstructured elements sufficiently small to model complex 3D subsurface structures and ensure numerical convergence across frequency ranges critical to structural response. Furthermore, accurately capturing actual nonlinear soil responses requires complex, high-fidelity physical models (constitutive laws), resulting in large computational costs. 
Prior research to reduce computational costs of 3D nonlinear ground amplification analyses include urban-scale dynamic simulations featured in SC Gordon Bell Prize sessions \cite{sc14, sc15}. However, these designs heavily prioritized time-to-solution based on simplified physical models, exhibiting limitations when optimizing energy-to-solution on modern architectures or adapting to the memory-intensive, high-fidelity constitutive models addressed in this study. 
Therefore, developing a new set of core kernels optimized for modern heterogeneous computing environments is essential to realizing the proposed workflow.

\subsection{Mathematical Formulation and Computational Bottlenecks}
We describe the 3D ground amplification analysis based on \cite{asme2014}, which is used as a basis of \cite{sc14, sc15}.
Owing to its capability in modeling complex geometries and naturally satisfying stress-free boundary conditions at the ground surface, the nonlinear wave equation---with material properties that evolve over time according to the physical soil model---is discretized via the finite element method \cite{FEM}. The Newmark-$\beta$ method \cite{newmark} is used for time integration, leading to solving the following equation at each time step to evaluate the response of the nonlinear time-evolution problem:
\begin{eqnarray}
\left( \frac{4}{dt^2}\mathbf{M}+\frac{2}{dt}\mathbf{C}^n+\mathbf{K}^n\right)\delta \mathbf{u}^n = \nonumber \\
\mathbf{f}^n-\mathbf{q}^{n-1}+\mathbf{C}^n\mathbf{v}^{n-1}+\mathbf{M}\left(\mathbf{a}^{n-1}+\frac{4}{dt}\mathbf{v}^{n-1}\right),
\label{eq:GE}
\end{eqnarray}
with
\begin{equation}
\begin{cases}
\mathbf{q}^n=\mathbf{q}^{n-1}+\mathbf{K}^n\delta\mathbf{u}^n,\\
\mathbf{u}^n=\mathbf{u}^{n-1}+\delta\mathbf{u}^n, \\
\mathbf{v}^n=-\mathbf{v}^{n-1}+\frac{2}{dt}\delta\mathbf{u}^n, \\
\mathbf{a}^n=-\mathbf{a}^{n-1}-\frac{4}{dt}\mathbf{v}^{n-1} +\frac{4}{dt^2}\delta\mathbf{u}^n.
\label{eq:newmarkbeta}
\end{cases}
\end{equation}
Here, $\mathbf{M}$, $\mathbf{C}^n$, and $\mathbf{K}^n$ denote the mass, damping, and stiffness matrices at the $n$-th time step, respectively.
Vectors $\delta\mathbf{u}^n$, $\mathbf{u}^n$, $\mathbf{v}^n$, $\mathbf{a}^n$, $\mathbf{f}^n$, and $\mathbf{q}^n$ represent the nodal vectors of displacement increment, displacement, velocity, acceleration, outer force, and inner force respectively, while $dt$ denotes the time step.
Rayleigh damping is employed, where the element damping matrix $\mathbf{C}_e^n$ is expressed using the element mass matrix $\mathbf{M}_e$ and element stiffness matrix $\mathbf{K}_e^n$ as $\mathbf{C}_e^n=\alpha \mathbf{M}_e+\beta \mathbf{K}_e^n$.
Here, coefficients $\alpha$ and $\beta$ are obtained by solving the following least-squares problem:
\[\text{minimize}\left[ \int_{f_{\min}}^{f_{\max}} \left(h^n -\frac12\left(\frac{\alpha}{2\pi f}+2 \pi f \beta\right) \right)^2 \text{d}f  \right], \]
where $f_{\max}$ and $f_{\min}$ are the upper and lower target frequencies, respectively, and $h^n$ is the damping ratio that evolves according to the physical soil model.
Consequently, Eq.~\eqref{eq:GE} requires solving a system of equations involving matrices $\mathbf{K}^n$ and $\mathbf{C}^n$, which change at each time step $n$ due to the physical nonlinear soil model.
Second-order tetrahedral elements are used as they are suitable for modeling complex geometry and due to the necessity in evaluating strain for the physical model.
Semi-infinite absorbing boundary conditions are applied to the bottom and side boundaries.

In contrast to the conventional simplified soil constitutive models used in \cite{sc14, sc15, asme2014}, this study adopts the multi-spring model \cite{multispring}---a high-fidelity physical constitutive law---to evaluate complex soil nonlinear responses more faithfully (see Appendix for details).
While this significantly enhances physical fidelity, it requires storing and updating a vast number of history data for each element, leading to a memory-intensive and computationally expensive analysis with stringent requirements on memory bandwidth and capacity.
Therefore, designing algorithms tailored to modern architectures, as described in the subsequent sections, becomes important.

\subsection{CPUGPU-Harvest: Energy-Aware Kernel for Resource Harvesting}
A standard baseline implementation for this core kernel would perform multi-node GPU computations where history data for the high-fidelity physical model are stored in GPU memory, the target matrix is stored in $3 \times 3$ block CRS format, and the system is solved using the Conjugate Gradient method with a $3 \times 3$ block Jacobi preconditioner (hereafter referred to as \texttt{CRSGPU}; see Algorithm~\ref{alg:CRSGPU}). 
On the other hand, during multi-case runs aimed at generating datasets for NNs, maximizing energy efficiency and total system throughput across the supercomputer takes precedence over minimizing the execution time of a single case.
Therefore, extending \cite{ICCS26}, we developed \texttt{CPUGPU-Harvest}, a multi-node dense kernel that exhibits superior energy-to-solution, leverages CPU resources by taking advantage of recent improvement in CPU-GPU data transfer bandwidths (e.g., \cite{PCIe, GH200}), and enables coupled CPU-GPU execution that fits into idle system backfill slots, executing multi-case analyses with minimal resource overhead.
Here, \texttt{CPUGPU-Harvest} is expected to be a broadly applicable approach for conducting many cases of memory-intensive simulations, such as those employing high-fidelity constitutive laws.

\begin{algorithm}[tb]
\caption{Baseline method: \texttt{CRSGPU}. $\mathbf{D}$ indicates the stiffness matrix evaluated by the multispring method, while $\theta$ indicates the nonlinear spring parameters. CRS-PCG indicates $3\times3$ block Jacobi preconditioned conjugate gradient solver with CRS-based matrix-vector products. $\mathbf{A}$ and $\mathbf{b}$ indicate left hand side matrix and right hand side vector of Eq.~\eqref{eq:GE}, respectively. All computation is done in FP64.}
\label{alg:CRSGPU}
\begin{algorithmic}[1]
\small{
\FOR{$n = 1;~ n \le n_t;~ n=n+1$}
\STATE ~~ $\delta \mathbf{u}^{n} \Leftarrow$ CRS-PCG($\mathbf{A}$, $\mathbf{b}^{n}$)@GPU \\
\STATE ~~ $\{ \mathbf{D}^{n}, \theta^{n}\} \Leftarrow$ Multispring($\delta \mathbf{u}^{n}, \theta^{n-1}$)@GPU \\
\STATE ~~ $\mathbf{A} \Leftarrow$ UpdateCRS($\mathbf{D}^{n}$)@GPU \\
\ENDFOR
}
\end{algorithmic}
\end{algorithm}

Our point of departure, the method in \cite{ICCS26}, was specifically designed for a single GH200 node environment equipped with a CPU with large memory and a high-speed GPU (i.e., configuration with a 72-core Grace CPU with 480 GB of memory and an H100 GPU with 96 GB of memory connected via a 900 GB/s NVLink-C2C interconnect), posing challenges in application to general heterogeneous or large-scale environments.
In this study, we generalize this memory/compute decoupling scheme for heterogeneous environments with an arbitrary number of nodes, while integrating the Element-by-Element (EBE) method \cite{EBE} and multigrid preconditioning to achieve high versatility and throughput via concurrent multi-case execution.

Task parallel and task splitting approaches are two primary methods for utilizing both compute and memory resources across CPUs and GPUs.
In the task parallel approach, dependencies between computational kernels are analyzed to assign independent kernels across CPUs and GPUs, enabling parallel utilization of CPU and GPU resources (e.g., \cite{Augonnet2011}). In contrast, the task splitting approach divides the domain within each kernel into subdomains handled by the CPU and GPU, computing each kernel in parallel (e.g., \cite{Pearce2018}). See e.g., \cite{Mital2015,Raju2018} for related studies on cooperative CPU-GPU computing.
In both approaches, tasks are assigned to the CPU or GPU with the compute and memory in a set; consequently, depending on the system, either compute performance or memory capacity becomes a bottleneck, preventing fully effective resource utilization. 
To address this, \cite{ICCS26} decoupled memory and compute, storing large history data in CPU memory while executing computations on high-performance GPUs, thereby leveraging the strengths of both CPU memory and high-speed GPUs (specifically, multi-spring calculations are executed on the GPU while asynchronously sending/receiving history data stored in CPU memory, enabling large-scale computations that exceed GPU memory capacity). 
Although the method in \cite{ICCS26} targeted only systems where CPU memory was significantly larger than GPU memory, this study generalizes the memory/compute decoupling technique to accommodate architectures with arbitrary CPU-to-GPU memory capacity ratios (Algorithm~\ref{alg:CPUGPU-Harvest}).
Here, a portion of the history data is stored in CPU memory and the remaining portion in GPU memory, with an adjustable allocation ratio that enables system-wide memory resource utilization regardless of the system's CPU-to-GPU memory ratio.
History data stored in CPU memory is updated by the aforementioned pipelined transfer and GPU computation, whereas that stored in GPU memory is directly updated by the GPU.

\begin{algorithm}[tbp]
\caption{\texttt{CPUGPU-Harvest}. $\{\theta^{n}_{part(i)}\}$ indicates parts of the spring parameters allocated on GPU memory ($i=0$) or CPU memory ($i=1,2,...,n_p$). Lines 14 and 15 (or lines 17 and 18) are conducted asynchronously such that the CPU-GPU transfers and GPU computation are overlapped. "que" stores the index of multispring data on CPU (excluding those residing on GPUbuffer1 and 2. For example, in the initial step of $n=1$, que = $\{3,4,...,n_p\}$). EBE-MultiIPCG is a conjugate gradient solver with multi-grid and mixed-precision preconditioner, with element-by-element method used for conducting matrix-vector products. $m$ cases are solved together to reduce random accesses. Computation in preconditioner of EBE-MultiIPCG is done in FP32, while all other computation is done in FP64.}
\label{alg:CPUGPU-Harvest}
\begin{algorithmic}[1]
\small{
\STATE {\bf // Copy part of CPU data to GPU buffers}
\STATE CPU[$\theta^{0}_{part(1)}$]$\rightarrow$ GPUbuffer1 \\
\STATE CPU[$\theta^{0}_{part(2)}$]$\rightarrow$ GPUbuffer2 \\
\FOR{$n = 1;~ n \le n_t;~ n=n+1$}
\STATE {\bf // Solver}
\STATE $\delta \mathbf{u}^{n} \Leftarrow$ EBE-MultiIPCG($\mathbf{D}^{n-1}$, $\mathbf{b}^{n}$)@GPU \\
\STATE {\bf // Multispring on GPU memory}
\STATE $\{ \mathbf{D}^{n}_{part(0)}, \theta^{n}_{part(0)}\} \Leftarrow$ Multispring($\delta \mathbf{u}^{n}, \theta^{n-1}_{part(0)}$)@GPU \\
\STATE {\bf // Startup of piplined multispring on CPU memory}
\STATE Compute MS on GPUbuffer1; GPUbuffer1.state $\Leftarrow$ done\\
\STATE {\bf // Piplined multispring on CPU memory} \\
\WHILE{que.size() != 0}
\IF{GPUbuffer1.state == done}
\STATE GPUbuffer1$\rightarrow$CPU; CPU[$\theta^{0}_{part(\text{que.pop()})}$]$\rightarrow$GPUbuffer1 \\
\STATE Compute MS on GPUbuffer2; GPUbuffer2.state $\Leftarrow$ done \\
\ELSE
\STATE GPUbuffer2$\rightarrow$CPU; CPU[$\theta^{0}_{part(\text{que.pop()})}$]$\rightarrow$GPUbuffer2 \\
\STATE Compute MS on GPUbuffer1; GPUbuffer1.state $\Leftarrow$ done \\
\ENDIF \\
\ENDWHILE \\
\STATE {\bf // Closing of piplined multispring on CPU memory}
\IF{GPUbuffer1.state == done}
\STATE Compute MS on GPUbuffer2; GPUbuffer2.state $\Leftarrow$ done \\
\ELSE
\STATE Compute MS on GPUbuffer1; GPUbuffer1.state $\Leftarrow$ done \\
\ENDIF \\
\ENDFOR
}
\end{algorithmic}
\end{algorithm}

Furthermore, for sparse matrix-vector multiplication in the solver, we adopt the EBE method, which generates element matrices on the fly and multiplies them with the right-hand-side vector rather than reading the global matrix from memory.
Although this increases floating-point operations compared to CRS-based sparse matrix-vector multiplication, reducing the data volume read from memory yields speedup on modern GPUs. Moreover, it eliminates the need to update the CRS matrix at every time step (Algorithm~\ref{alg:CRSGPU}, Line 4), providing additional speedups. 
In addition, eliminating the need to store the global matrix in CRS format frees up memory capacity, allowing multiple analysis cases ($m$ cases) with varying input ground motions to be computed concurrently; this reduces random data accesses during sparse matrix-vector multiplication, leading to further performance gains. 
Additionally, using multigrid preconditioning in the solver enables efficient solution convergence. 
By incorporating halo exchanges into sparse matrix-vector multiplication and \texttt{MPI\_Allreduce} into inner products within the solver, parallel execution across multiple computer nodes is achieved.

The developed kernel, \texttt{CPUGPU-Harvest}, keeps high-speed GPUs continuously active while fully utilizing both CPU and GPU memory capacities. 
Executing such dense computations on a small number of nodes minimizes data access and communication overheads, enabling the efficient harvesting of idle supercomputer capabilities with high energy efficiency.

\subsection{EBEGPU-Urgent: Time-to-Solution-Optimized GPU Kernel}
During the post-disaster emergency phase, minimizing the latency and time-to-solution for a single analysis case becomes top priority, where slight degradation in energy efficiency can be tolerated.
Therefore, by extending \texttt{CRSGPU}, we developed \texttt{EBEGPU-Urgent}, a pure GPU kernel designed to enable urgent decision-making within 30 minutes.
The specific algorithm of \texttt{EBEGPU-Urgent} (see Algorithm~\ref{alg:EBEGPU-Urgent}) targets a single analysis case and improves throughput by adopting the EBE formulation for sparse matrix-vector multiplication, which also eliminates the need for CRS matrix updates.
Furthermore, under the premise of deploying large node allocations, all history data are stored entirely in high-speed GPU memory, and all computations are executed on high-speed GPUs. 
Overlapping sparse matrix-vector multiplication computation with halo exchange communication further enhances scalability across large node counts.
Although CPU cores and memory are not utilized, employing a large number of high-speed GPUs achieves extreme computational speed.

\begin{algorithm}[tb]
\caption{{\texttt{EBEGPU-Urgent}. $\mathbf{D}$ indicates the stiffness matrix evaluated by the multispring method, while $\theta$ indicates the nonlinear spring parameters. EBE-PCG indicates a $3\times3$ block Jacobi preconditioned conjugate gradient solver with EBE-based matrix-vector products. All computation is done in FP64.}}
\label{alg:EBEGPU-Urgent}
\begin{algorithmic}[1]
\small{
\FOR{$n = 1;~ n \le n_t;~ n=n+1$}
\STATE ~~ $\delta \mathbf{u}^{n} \Leftarrow$ EBE-PCG($\mathbf{D}^{n-1}$, $\mathbf{b}^{n}$)@GPU \\
\STATE ~~ $\{ \mathbf{D}^{n}, \theta^{n}\} \Leftarrow$ Multispring($\delta \mathbf{u}^{n}, \theta^{n-1}$)@GPU \\
\ENDFOR
}
\end{algorithmic}
\end{algorithm}

\subsection{Performance Evaluation on the Miyabi Supercomputer}

The performance of \texttt{CPUGPU-Harvest} and \texttt{EBEGPU-Urgent} is evaluated in comparison with the baseline \texttt{CRSGPU} on Miyabi \cite{miyabi}. 
Miyabi is a jointly operated supercomputer system of the Joint Center for Advanced High Performance Computing (JCAHPC), managed by the Information Technology Center at The University of Tokyo and the Center for Computational Sciences at the University of Tsukuba. In this study, we utilize Miyabi-G, where each compute node is equipped with a single NVIDIA GH200 Grace Hopper Superchip.
Each node has one H100 GPU (96 GB, 4.0 TB/s) and one Grace CPU (120 GB, 512 GB/s) interconnected via NVLink-C2C (900 GB/s).
A total of 1,120 compute nodes are connected via InfiniBand NDR200 in a full-bisection Fat-Tree topology.
The implementations were performed at the same level of optimization as in \cite{WACCPD2024}, which conducts coupled CPU-GPU execution for similar analyses.
Measurements reported below were obtained across all time steps ($n_t=14500$), with energy consumption evaluated using module-wide power values---including the CPU, GPU, and memory subsystems---obtained via \texttt{nvidia-smi -q -d POWER}.

First, we compare the performance during resource harvesting in the pre-earthquake phase.
Here, throughput and energy-to-solution are compared when using 4 nodes, which is required by the conventional \texttt{CRSGPU} method to solve the problem within a practical execution time (Table~\ref{tb:performance_preeq}).
We first compare the baseline \texttt{CRSGPU} with \texttt{CPUGPU-Harvest} executing a single case ($m=1$).
In \texttt{CPUGPU-Harvest}, half of the multi-spring data is stored in CPU memory and the other half in GPU memory, executing with pipelined overlap between CPU data transfers and GPU computation.
Although pipeline execution overhead increases the execution time of the multi-spring (MS) section from 799 s to 879 s, the solver section is reduced from 5,352 s to 3,296 s owing to multigrid preconditioning and the EBE method. Combined with the elimination of CRS matrix updates, the overall application throughput improves by a factor of 2.1 (8825/4255).
At this stage, as shown in Table~\ref{tb:memory_preeq}, \texttt{CPUGPU-Harvest} ($m=1$) leaves room in both CPU and GPU memory, enabling concurrent execution of $m=4$ cases (the multi-spring data size is 20.4 GB per node per case, half of which is assigned to the CPU side and half to the GPU side even when $m=4$). 
This reduces random memory accesses, further improving the throughput (execution time per analysis case) of the solver section from 3,296 s to 1,890 s compared to $m=1$. For the entire application, this corresponds to an 1.8-fold throughput improvement relative to $m=1$, and a 3.7-fold improvement relative to the conventional \texttt{CRSGPU} method (8,825 s $\rightarrow$ 2,370 s/case).
Furthermore, using the EBE method reduces total memory transfers per case relative to \texttt{CRSGPU}, which lowers power consumption from 630 W to 569 W. Consequently, the energy-to-solution is reduced to 0.24 times that of the baseline (a 76\% reduction in energy consumption).
Note that executing \texttt{EBEGPU-Urgent} on 4 nodes (described below) requires 4,143 s (10.2 MJ), indicating that \texttt{CPUGPU-Harvest} with dense computations delivers superior performance in terms of both throughput and energy-to-solution. 
Thus, a high-throughput, energy-efficient algorithm tailored to the pre-earthquake phase has been successfully realized. Using this approach, the 100-case training dataset used in the application example in Section~\ref{sct4} can be computed in sets of 4 nodes for 2.6 hours ($2,370~\mathrm{s} \times 4$ cases), achieving a granularity that can be flexibly scheduled into short idle backfill slots during normal supercomputer operations.
Moreover, the entire dataset can be constructed in a practical resource size of 263 node-hours in total.

\begin{table}[tb]
{\small
\begin{center}
\caption{Performance for Pre-Earthquake Phase measured on 4 nodes. \texttt{CGH} stands for \texttt{CPUGPU-Harvest}. $m$ is the number of cases solved simultaneously. MS indicates time for multispring computation. Time is shown per case.}
\label{tb:performance_preeq}
\begin{tabular}{lccccc}
\hline
                  & Elapsed time in sec. & Power & Required  \\ 
                  & Total (Solver, MS, CRS) & /node & energy \\ \hline
\texttt{CRSGPU}               & 8825 (5352, 799, 2590) & 630 W & 22.2 MJ \\
\texttt{CGH} [$m=1$] & 4255 (3296, 879, -- )  & 504 W & 8.58 MJ \\
\texttt{CGH} [$m=4$] & 2370 (1890, 419, -- )  & 569 W & 5.40 MJ \\
\hline
\end{tabular}
\end{center}
}
\end{table}

\begin{table}[tb]
{\small
\begin{center}
\caption{Memory usage for Pre-Earthquake Phase measured on 4 nodes. Per node values obtained by \texttt{numactl -H} are shown.}
\label{tb:memory_preeq}
\begin{tabular}{lccccc}
\hline
                                & CPU memory & GPU memory \\ \hline
\texttt{CRSGPU}                 & 19.3 GB    & 43.1 GB \\
\texttt{CPUGPU-Harvest} [$m=1$] & 28.4 GB    & 23.4 GB \\
\texttt{CPUGPU-Harvest} [$m=4$] & 75.0 GB    & 89.0 GB \\
\hline
\end{tabular}
\end{center}
}
\end{table}

Next, to evaluate urgent computing performance during the post-earthquake phase, we measure the strong scaling performance of \texttt{EBEGPU-Urgent} (Table~\ref{tb:performance_posteq}). 
Although overhead from strong scaling causes a decrease in parallel efficiency, adhering to the emergency-phase design philosophy of prioritizing latency minimization enables the complete 14,500-step analysis to be computed in just 771 s (12.1 min) when using 128 nodes. 
The required energy on 128 nodes is 24.1 MJ; thus, \texttt{EBEGPU-Urgent} achieves an 11.4-fold speedup (8825/771) over \texttt{CRSGPU} on 4 nodes (22.2 MJ) while consuming roughly comparable energy, making it well-suited for immediate post-earthquake deployments where power consumption should be constrained. 
In this manner, an ultra-fast, single-case time-to-solution tailored to the post-earthquake phase is achieved with high energy efficiency.

\begin{table}[tb]
{\small
\begin{center}
\caption{Performance for Post-Earthquake Phase. MS indicates time for multispring computation.}
\label{tb:performance_posteq}
\begin{tabular}{lccccc}
\hline
                    & Compute  & Elapsed time                & Required \\ 
                    & nodes    & Total (Solver, MS) &   energy \\ \hline
                       & 4  & 4143 s (3295 s, 769 s) & 10.2 MJ \\
                       & 8  & 2321 s (1886 s, 389 s) & 10.7 MJ \\
\texttt{EBEGPU-} & 16 & 1411 s (1179 s, 202 s) & 10.7 MJ \\
\texttt{Urgent}  & 32 &  965 s (834 s,  111 s) & 12.6 MJ \\
                       & 64 &  855 s (774 s,   63 s) & 17.5 MJ \\
                       & 128 & 771 s (725 s,   31 s) & 24.1 MJ \\
\hline
\end{tabular}
\end{center}
}
\end{table}

\section{Validation and Application: High-Fidelity Urban Ground Motion Estimation}
\label{sct4}
In this section, we apply the proposed workflow to immediately estimate spatial ground motion distributions across the target region based on ground motion recorded at a single surface observation point.
During large earthquakes, ground motions are typically observed at only a sparse set of surface stations relative to the domain size and spatial resolution required for post-disaster damage estimation; thus, there is strong demand for methodologies capable of rapidly estimating spatial ground motion distributions as demonstrated in this study.

\subsection{Simulation Setup and High-Fidelity Geotechnical Model}
\begin{figure}[tb]
\begin{center}
\includegraphics[width=\hsize]{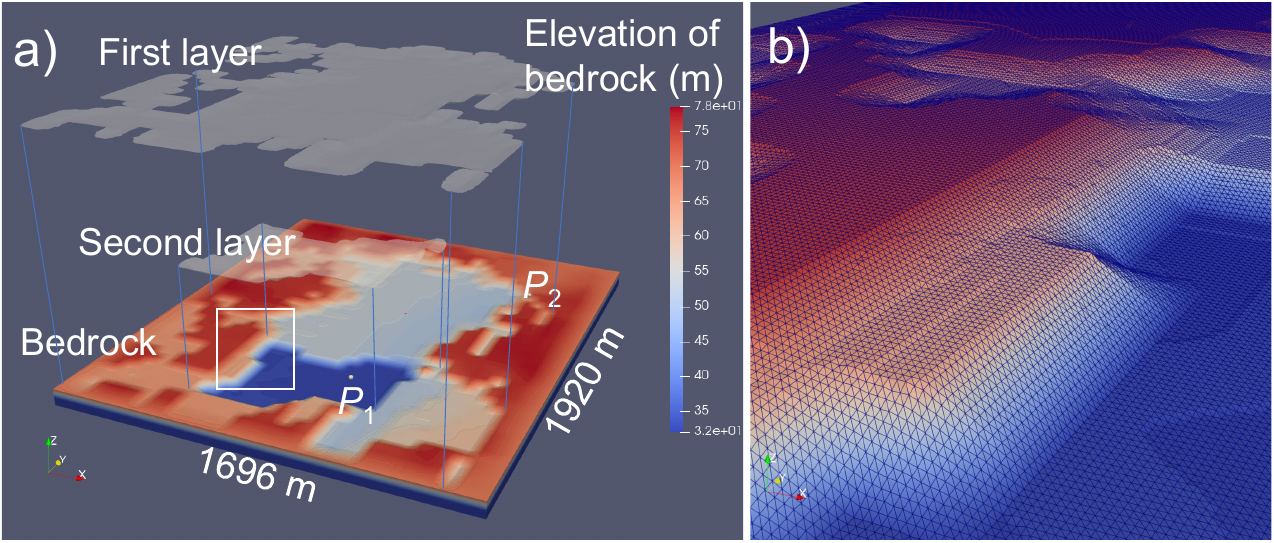}
\end{center}
\caption{(a) 3D ground structure model, and the position of observation points $P_1$ and $P_2$. (b) Close-up view of the region shown by the rectangle in (a).}
\label{fig:3Dmodel}
\end{figure}

As an example target site, we consider a location near Yokohama City, Kanagawa Prefecture, Japan. 
This site features soft sedimentary layers forming a complex 3D structure, resulting in significantly larger ground shaking compared to surrounding areas. 
That is, body waves are converted into surface waves and trapped, inducing large 3D nonlinear ground amplification, making it an ideal site for the evaluation conducted in this section considering damage-inducing 3D nonlinear ground amplification. 
To understand the ground motion mechanisms at this site and obtain insights for disaster mitigation, a detailed geotechnical structure model was constructed by a dedicated committee through comprehensive investigations using various datasets; thus, this study utilizes this ground model. 
The constructed model forms a three-layer stratified structure over an area of 1,696 m east-west by 1,920 m north-south, with soil material properties summarized in Table~\ref{tb:material}. 
In \cite{asme2014}, 3D nonlinear analysis results were compared against observation data recorded during the devastating 2011 Tohoku earthquake, demonstrating good validation performance of this model. 
For this model, similar to \cite{asme2014}, frequencies up to 2.5 Hz are targeted to capture the primary damage-inducing components, and a finite-element model using unstructured second-order tetrahedral elements was generated to ensure at least 10 elements (21 nodes) per wavelength (given a minimum shear wave velocity of 100 m/s and a wavelength of 40 m at 2.5 Hz, the minimum element size is set to approximately 4 m based on the 10-element threshold). 
Figure~\ref{fig:3Dmodel} shows the generated 3D ground finite-element model with complex geometry comprising three soil layers (32,502,492 degrees of freedom and 7,781,075 tetrahedral elements). 
Here, a Cartesian coordinate system is used with the origin placed at the southwest bottom corner of the model, where the $x$-, $y$-, and $z$-axes represent the east-west, north-south, and vertical directions, respectively. 
In the 3D nonlinear ground amplification analysis conducted in this study, the time step is set to $dt=0.005$ s as in \cite{asme2014}, with a relative error convergence threshold of $10^{-8}$, running for 14,500 time steps.

\begin{table}[tb]
{\small
\begin{center}
\caption{Material properties of the soil structure}
\label{tb:material}
\begin{tabular}{lccccc}
\hline
 & $V_p$ m/s & $V_s$ m/s & $\rho$ kg/m$^3$ & $h_{max}$ & $\gamma_r$ \\ \hline
First layer & 700 & 100 & 1500 & 0.23 & 0.007 \\
Second layer & 1400 & 300 & 1800 & 0.23 & 0.001 \\
Bedrock & 2100 & 700 & 2100 & 0.01 & $\infty$ \\
\hline
\end{tabular}
\end{center}
}
\end{table}

To evaluate the estimation performance of the proposed workflow under nonlinear behavior driven by realistic large earthquake motions, and considering the localized domain, uniform plane waves are input from the bottom of the ground model—a practice widely adopted in conventional studies. Specifically, the Kobe wave (hereafter referred to as true input bedrock motion $b_{\text{ref}}^{\text{K}}$) and the Chuetsu-oki wave (hereafter referred to as true input bedrock motion $b_{\text{ref}}^{\text{C}}$) are applied as inputs, considering cases where ground motions are observed at points $P_1$ and $P_2$ shown in Fig.~\ref{fig:3Dmodel}. That is, we rapidly estimate spatial ground motion distributions using solely the surface waveform observed at $P_2$, and validate the estimation accuracy against the waveform observed at $P_1$.
Note that $b_{\text{ref}}^{\text{K}}$ was generated based on seismic motion recorded at Nakayamate, Chuo-ku, Kobe City, Hyogo Prefecture during the devastating 1995 Kobe earthquake (Hyogoken-Nanbu earthquake) \cite{eqdata}, which is widely used in seismic design in Japan.
Meanwhile, $b_{\text{ref}}^{\text{C}}$ was generated based on seismic motion recorded at Komeda, Izumozaki Town, Niigata Prefecture during the severe 2007 Niigata-ken Chuetsu-oki earthquake \cite{eqdata}. 
Specifically, because these records are surface observation data, their amplitudes were halved to convert them into bedrock input motions. Furthermore, following \cite{asme2014}, frequency components up to 2.5 Hz were extracted to focus on the frequency band associated with heavy damage (i.e., a 0.2–0.5–2.4–2.5 Hz bandpass filter is applied).
$b_{\text{ref}}^{\text{C}}$ is characterized by having slightly higher frequency content compared to $b_{\text{ref}}^{\text{K}}$.
Since the waveform evaluation results were similar across $x$-, $y$-, and $z$-components, the $x$-component is presented as a representative case.
Additionally, the refinement process is configured to run whenever the relative $L_1$ error (defined as $Err$ in Table~\ref{tb:accuracy}) is equal to or exceeds 0.10.

\begin{table}[tb]
\caption{Quantitative accuracy of estimated wave ($x$-component of input wave and wave at $P_1$). Here, $Err(x, y) = \sum_{i=1}^{n_t}|x_{i} - y_{i}| / \sum_{i=1}^{n_t}|y_{i}| $, where $i$ is the time step number and $n_t$ is the number of time steps.}
\label{tb:accuracy}
\centering
\textbf{\small (a) Kobe scenario}\\
\begin{tabular}{|c|c|c|}
\hline
Phase I: & $ Err(s_{\text{sim}, P_1}^{\text{K}, (1)}, s_{\text{ref}, P_1}^{\text{K}})$: 0.128 & $ Err(b_{\text{est}}^{\text{K}, (1)}, b_{\text{ref}}^{\text{K}})$: 0.110 \\
\hline
Phase II: & $ Err(s_{\text{sim}, P_1}^{\text{K}, (2)}, s_{\text{ref}, P_1}^{\text{K}})$: 0.073 & $ Err(b_{\text{est}}^{\text{K}, (2)}, b_{\text{ref}}^{\text{K}})$: 0.066 \\
\hline
\end{tabular}\\
\hspace{0.5 cm}\\
\textbf{\small (b) Chuetsu-oki scenario}\\
\begin{tabular}{|c|c|c|}
\hline
Phase I: & $ Err(s_{\text{sim}, P_1}^{\text{C}, (1)}, s_{\text{ref}, P_1}^{\text{C}})$: 0.119 & $ Err(b_{\text{est}}^{\text{C}, (1)}, b_{\text{ref}}^{\text{C}})$: 0.113 \\ 
\hline
Phase II: & $ Err(s_{\text{sim}, P_1}^{\text{C}, (2)}, s_{\text{ref}, P_1}^{\text{C}})$: 0.084 & $ Err(b_{\text{est}}^{\text{C}, (2)}, b_{\text{ref}}^{\text{C}})$: 0.081 \\
\hline
\end{tabular}
\end{table}

\subsection{Case I: Kobe Scenario}

\begin{figure*}[tb]
\begin{center}
\includegraphics[width=\hsize]{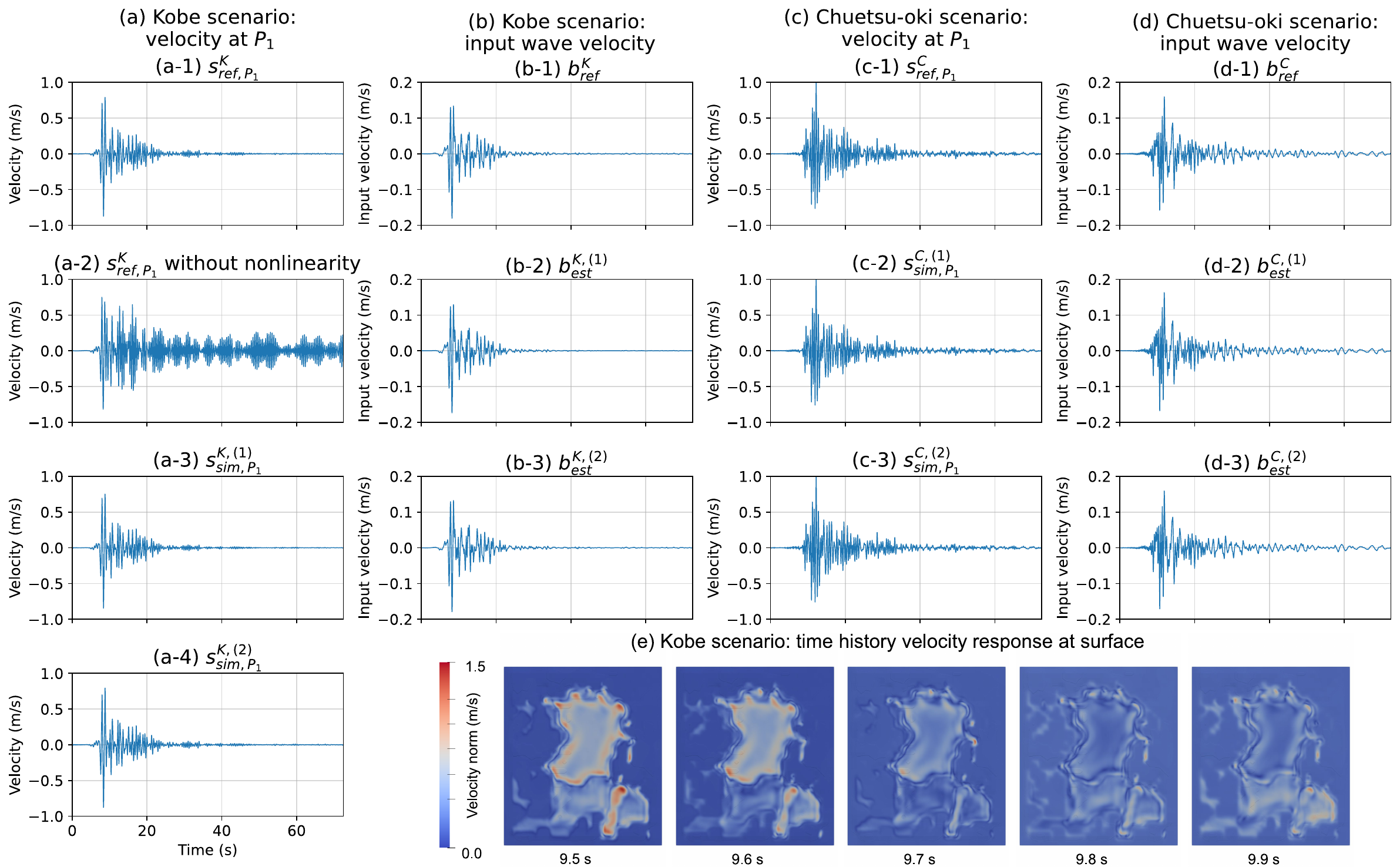}
\end{center}
\caption{Results for Kobe and Chuetsu-oki scenarios. Kobe scenario: (a) waves at $P_1$ , (b) input wave, and (e) time history response distribution. Chuetsu-oki scenario: (c) waves at $P_1$ , (d) input wave. The $x$-component is visualized for the waveforms.
}
\label{fig:response}
\end{figure*}

First, $b_{\text{ref}}^{\text{K}}$ was input into the ground model, and a 3D ground amplification analysis using the high-fidelity physical model described in the previous section was conducted (hereafter considered as the reference solution). 
Figure~\ref{fig:response}(e) shows the time-history velocity norm at the surface $s_{\text{ref}}^{\text{K}}$.
We can see that complex 3D nonlinear responses reflecting the complex ground structure occur, exhibiting significant time-history variations and vastly different responses depending on the location.
From this analysis result, surface waveforms ($s_{\text{ref}, P_1}^{\text{K}}$ and $s_{\text{ref}, P_2}^{\text{K}}$) were observed at points $P_1$ and $P_2$ shown in Fig.~\ref{fig:3Dmodel} 
(Fig.~\ref{fig:response}(a-1) shows the waveform at $P_1$).

To examine the effects of the high-fidelity physical model, waveforms computed without enabling the physical model were also observed 
(Fig.~\ref{fig:response}(a-2) shows the waveform at $P_1$).
Specifically, because this physical model induces nonlinearity---resulting in stiffness degradation and damping increase---based on the reference strain $\gamma_r$, setting a large value for $\gamma_r$ prevents the physical model from exhibiting nonlinear effects.
Therefore, $\gamma_r = 10$ was set here to suppress nonlinearity.
In the linear computation result, waves are trapped by the ground structure without undergoing stiffness degradation or damping, leading to significantly severe oscillations.
Conversely, under the nonlinear setting, substantial stiffness reduction and damping lead to markedly different waveforms, highlighting the importance of employing a high-fidelity physical model.

\textbf{Phase I: Rapid First-Order Estimation:} We attempt immediate ground motion distribution estimation using the proposed workflow.
First, we attempt to inversely estimate the input bedrock wave from $s_{\text{ref}, P_2}^{\text{K}}$.
Although this process is a nonlinear optimization problem to which various methods can be applied, we employ rapid evaluation via a NN surrogate model to achieve immediate post-disaster damage estimation.
That is, a NN that estimates input motions at the bottom of the model from surface observation data is constructed and used to infer the input wave.
Specifically, 100 random waves were generated with frequency components above 2.5 Hz filtered out and amplitudes uniformly distributed between -0.6 and 0.6 for the $x$- and $y$-components and between -0.3 and 0.3 for the $z$-component (this configuration reflects the fact that vertical components are typically smaller than horizontal components in actual earthquake ground motions).
Using these waves as inputs, 3D ground amplification analyses were performed using \texttt{CPUGPU-Harvest}, and response waveforms were recorded at $P_2$.
Through this procedure, 100 pairs of time-history data corresponding to input random waves and their surface responses at $P_2$ were obtained.
Using these data pairs, a NN surrogate model was constructed prior to the event using the method described in the Appendix; this model is denoted as $NN^{P_2}$.
Taking $s_{\text{ref}, P_2}^{\text{K}}$ as input, the initial bedrock wave inferred by $NN^{P_2}$ is denoted as $b_{\text{est}}^{\text{K}, (1)}$
(see Fig.~\ref{fig:response}(b-2)).
Although point $P_2$ is subject to complex subsurface structures and strong nonlinearities, $NN^{P_2}$---which accounts for these effects---infers $b_{\text{est}}^{\text{K}, (1)}$ with almost the same envelope and phase properties as that of $b_{\text{ref}}^{\text{K}}$, leading to low error of $Err = 0.110$
(see Figs.~\ref{fig:response}(b-1), \ref{fig:response}(b-2) and Table~\ref{tb:accuracy}).
Next, a 3D ground amplification analysis was executed using the estimated input $b_{\text{est}}^{\text{K}, (1)}$.
Comparing the computed waveform $s_{\text{sim}, P_1}^{\text{K}, (1)}$ at the unlearned validation point $P_1$ against the reference solution $s_{\text{ref}, P_1}^{\text{K}}$ yields $Err = 0.128$, also demonstrating close agreement 
(see Figs.~\ref{fig:response}(a-1), \ref{fig:response}(a-3) and Table~\ref{tb:accuracy}).

\textbf{Phase II: Interactive Refined Update:} Next, we attempt to improve the accuracy of the estimated spatial ground motion distribution.
Specifically, newly acquired data (the simulated surface waveform $s_{\text{sim}, P_2}^{\text{K}, (1)}$ at $P_2$ and $b_{\text{est}}^{\text{K}, (1)}$) are added to fine-tune $NN^{P_2}$ using the method detailed in the Appendix---adapting the model to observational facts while preserving the search capability derived from the random wave dataset---thereby constructing $NN^{P_2}_{\text{K}, (1)}$.
Using $NN^{P_2}_{\text{K}, (1)}$ and $s_{\text{ref}, P_2}^{\text{K}}$, an updated bedrock motion $b_{\text{est}}^{\text{K}, (2)}$ with improved accuracy is inversely estimated and used as input to execute a second 3D ground amplification analysis.
The bedrock motion $b_{\text{est}}^{\text{K}, (2)}$ and the resulting surface waveform $s_{\text{sim}, P_1}^{\text{K}, (2)}$ obtained at $P_1$ from this second analysis are shown in Figs.~\ref{fig:response}(b-3) and \ref{fig:response}(a-4).
Through this refinement process, the estimation error of the input bedrock motion decreases from 0.110 to 0.066, and the surface response error at the unlearned point $P_1$ also decreases from 0.128 to 0.073, clearly demonstrating an improvement in accuracy (see Table~\ref{tb:accuracy}).

\begin{figure}[tbp]
\begin{center}
\includegraphics[width=\hsize]{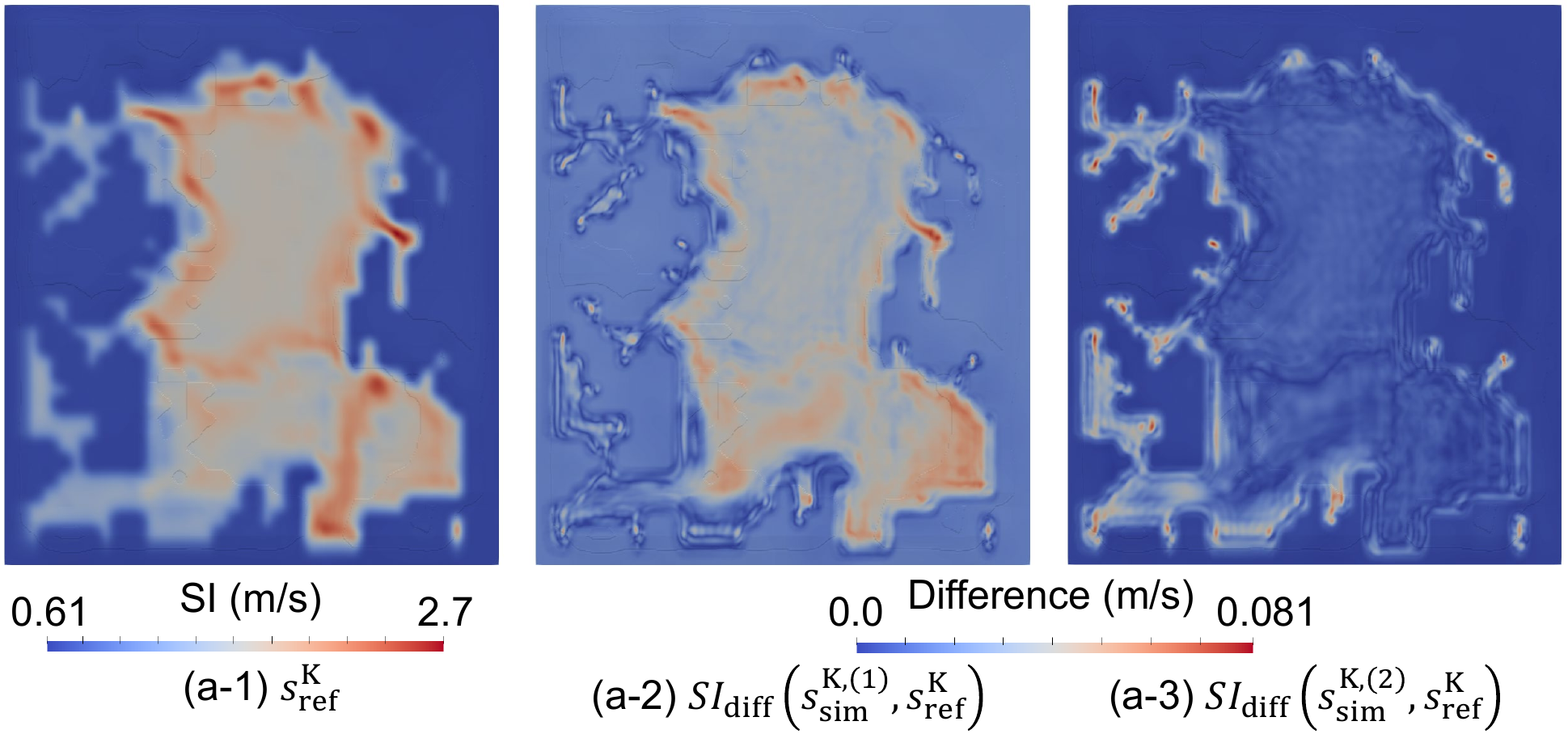}\\
{\small \textbf{(a) Kobe scenario}}\\
\includegraphics[width=\hsize]{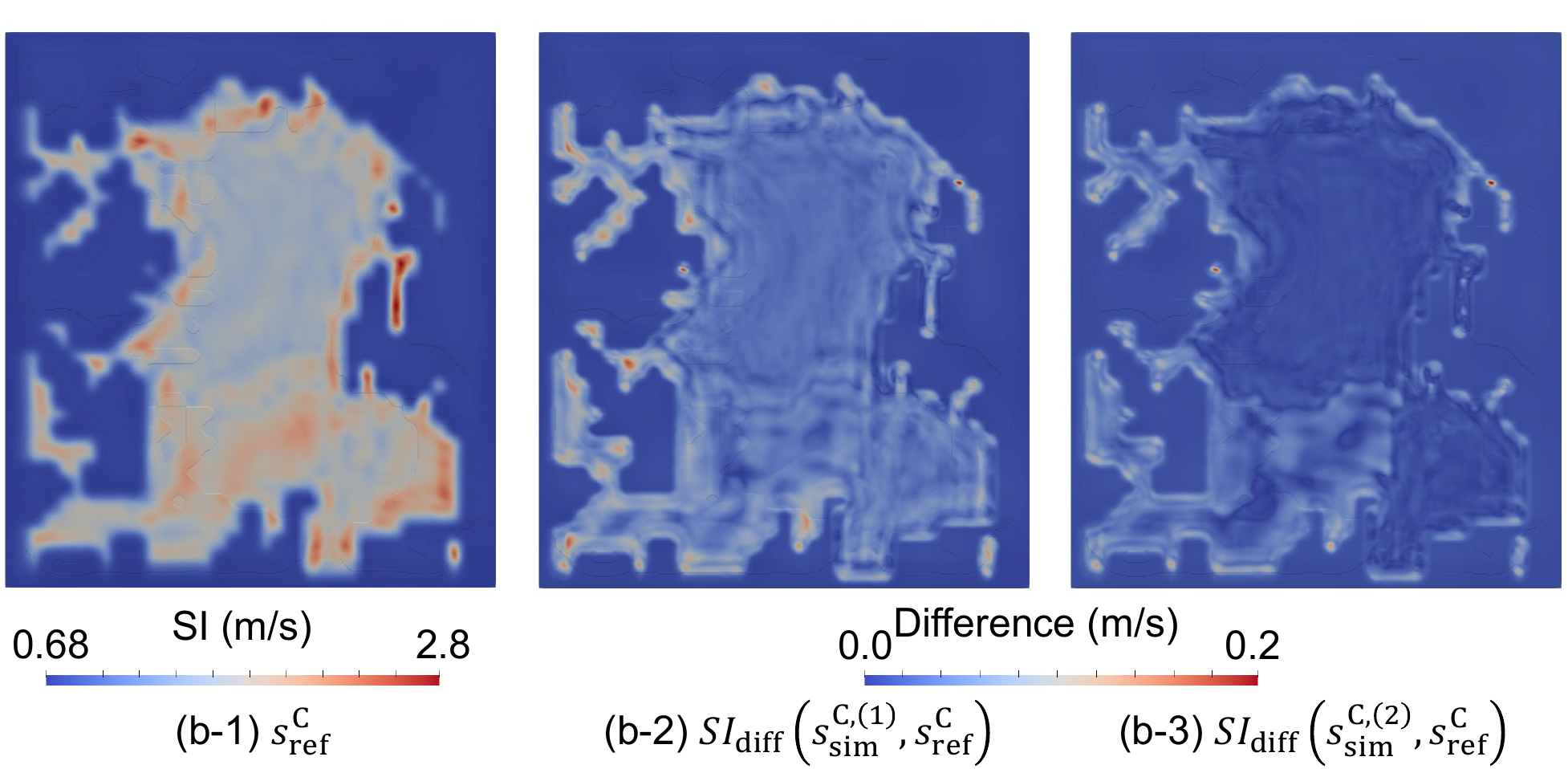}
{\small \textbf{(b) Chuetsu-oki scenario}}\\
\end{center}
\caption{SI values obtained at surface. 
Here, differences of SI values are computed as $SI_\text{diff}(\text{sim},\text{ref}) = \sqrt{\{SI_x(\text{sim})-SI_x(\text{ref})\}^2+\{SI_y(\text{sim})-SI_y(\text{ref})\}^2} $. }
\label{fig:SI}
\end{figure}

Finally, Fig.~\ref{fig:SI}(a) shows the surface Spectrum Intensity (SI) \cite{SIvalue} distributions obtained when applying $b_{\text{ref}}^{\text{K}}$, $b_{\text{est}}^{\text{K}, (1)}$, and $b_{\text{est}}^{\text{K}, (2)}$ as inputs (the SI value is a common measure used to estimate seismic damage to structures).
We can see that the 1st-order estimation gives SI distribution in high accuracy of 0.081/2.7 = 3\%, and that the refined update improves the SI prediction in regions with large SI values, which are of particular interest in disaster mitigation.
These comparisons indicate that the SI distribution---which directly correlates with damage assessment---can be estimated accurately, and that the refinement works effectively.

\subsection{Case II: Chuetsu-oki Scenario}

Similar to Case I, $b_{\text{ref}}^{\text{C}}$ was first input into the ground model to perform a 3D nonlinear ground amplification analysis, and the surface reference waveforms $s_{\text{ref}, P_1}^{\text{C}}$ and $s_{\text{ref}, P_2}^{\text{C}}$ at $P_1$ and $P_2$ were recorded.
Because the Chuetsu-oki wave and Kobe wave possess different characteristics, $b_{\text{ref}}^{\text{K}}$ and $b_{\text{ref}}^{\text{C}}$ are fundamentally distinct
(see Figs.~\ref{fig:response}(b-1) and \ref{fig:response}(d-1)).
Moreover, even when observed at the same point $P_1$, waveforms such as $s_{\text{ref}, P_1}^{\text{K}}$ and $s_{\text{ref}, P_1}^{\text{C}}$ exhibit markedly different properties, highlighting the importance of observing ground motions and conducting rapid post-event damage estimation that accounts for 3D nonlinear responses on an event-by-event basis 
(see Figs.~\ref{fig:response}(a-1) and \ref{fig:response}(c-1)).

\textbf{Phase I: Rapid First-Order Estimation:} We attempt immediate ground motion distribution estimation using the proposed workflow.
First, we attempt to inversely estimate the input bedrock wave from $s_{\text{ref}, P_2}^{\text{C}}$.
Inputting $s_{\text{ref}, P_2}^{\text{C}}$ into the pre-constructed shared surrogate model $NN^{P_2}$ described in Case I, we inferred the bedrock input wave $b_{\text{est}}^{\text{C}, (1)}$.
As shown in Figs.~\ref{fig:response}(d-1) and \ref{fig:response}(d-2),
despite having distinct characteristics from the Kobe wave---specifically having larger high-frequency components---it is accurately estimated by the pre-constructed surrogate model trained on random waves ($Err=0.113$).
Using the estimated input $b_{\text{est}}^{\text{C}, (1)}$, a 3D ground amplification analysis was executed.
The computed waveform $s_{\text{sim}, P_1}^{\text{C}, (1)}$ at the unlearned observation point $P_1$ also shows good agreement 
(Figs.~\ref{fig:response}(c-1) and \ref{fig:response}(c-2), $Err=0.119$ in Table~\ref{tb:accuracy}).

\textbf{Phase II: Interactive Refined Update:} Next, we attempt to improve the accuracy of the estimated spatial ground motion distribution from the first-order estimation.
Specifically, using the pair consisting of $b_{\text{est}}^{\text{C}, (1)}$ and the reproduced surface wave $s_{\text{sim}, P_2}^{\text{C}, (1)}$ from Phase I, $NN^{P_2}$ is fine-tuned on the fly using the method detailed in the Appendix---adapting the model to observation data while preserving search capability derived from random wave data---to rapidly construct an updated $NN^{P_2}_{\text{C}, (1)}$.
Using this updated model, the input bedrock wave $b_{\text{est}}^{\text{C}, (2)}$ is inversely estimated from the observed wave $s_{\text{ref}, P_2}^{\text{C}}$, and used as input to perform the second 3D nonlinear ground amplification analysis.
The resulting simulated waveform obtained at $P_1$ from this 3D analysis is denoted as $s_{\text{sim}, P_1}^{\text{C}, (2)}$.
Comparing these waveforms reveals that discrepancies in the first-order estimation are refined for both the input wave and the observed waveform at $P_1$, resulting in improved accuracy 
(see Figs.~\ref{fig:response}(d-3) and \ref{fig:response}(c-3);
the input bedrock wave error decreases from 0.113 to 0.081, and the waveform error at $P_1$ decreases from 0.119 to 0.084; see Table~\ref{tb:accuracy}).

Finally, Fig.~\ref{fig:SI}(b) shows the surface SI value distributions obtained when $b_{\text{ref}}^{\text{C}}$, $b_{\text{est}}^{\text{C}, (1)}$, and $b_{\text{est}}^{\text{C}, (2)}$ are used as inputs.
We can see that the 1st-order estimation gives SI distribution in high accuracy of 0.2/2.8 = 7\%, and that the refined update improves the SI prediction to 3\%.
These comparisons demonstrate that ground motion distributions can be accurately estimated even for complex seismic waves with distinct properties (Kobe and Chuetsu-oki cases), and that the interactive refinement functions effectively for both cases.

\subsection{Operational Timeline Analysis and Decision-Making Viability}

Finally, we quantitatively evaluate the time-to-insight and computational resource savings from the moment a ground motion is observed at point $P_2$ until decision-makers acquire the spatial time-history ground motion distribution for the Chuetsu-oki scenario.
The actual post-disaster operational timeline progresses through the following steps (in practice, these steps are executed concurrently in parallel across numerous target locations):
\begin{enumerate}
\item \textbf{Initial Surrogate Inference (1--2 s):} Input observation data $s_{\text{ref}, P_2}^\text{C}$ into the preconstructed $NN^{P_2}$ to infer the bedrock wave $b_{\text{est}}^{\text{C}, (1)}$.
\item \textbf{1st-Order Physical Simulation (12.1 min):} Using the estimated bedrock wave, run a 3D nonlinear analysis on 128 nodes (128 GPUs) of Miyabi-G via the \texttt{EBEGPU-Urgent} kernel. A physically consistent first-order spatial time-history distribution is obtained approximately 12 minutes post-event (completing immediate assessment).
\item \textbf{On-the-fly Fine-Tuning (1.1 min):} Based on the discrepancy between observed waves and the surface waves obtained from the 1st-order forward analysis, construct $NN^{P_2}_{\text{C}, (1)}$ via dynamic incremental training on 1 node (1 GPU) of Miyabi-G.
\item \textbf{Interactive Refined Simulation (12.1 min):} Execute a second 3D nonlinear analysis using the re-inferred bedrock wave $b_{\text{est}}^{\text{C}, (2)}$.
\end{enumerate}

Across the entire process above (1st-Order + Refinement), the pure computational time-to-insight is 25.3 minutes, and computational resource consumption is limited to 51.7 node-hours (128 nodes $\times$ 24.2 min $+$ 1 node $\times$ 1.1 min) when refinement is triggered by automated thresholding.
In conventional methods, achieving a comparable level of accuracy would require at least $10^{2-3}$ forward analysis runs (approximately 20.2--202 hours on 128 nodes or 2,580--25,800 node-hours), even if utilizing the fast \texttt{EBEGPU-Urgent} kernel developed in this study, rendering real-time emergency operations physically impossible.
The proposed workflow cuts required computation time and resource consumption by over 97.9\% compared to conventional approaches while keeping physics-based simulations at its core.
Consequently, it delivers 1st-order results just 12.1 minutes post-event while reliably satisfying the "30-minute" urgent decision-making time frame, even when performing interactive refinement in response to decision-makers' requirements.

\section{Conclusion}
\label{sct5}

In this study, we proposed a computational workflow for immediate post-disaster damage estimation that couples high-fidelity 3D nonlinear physics simulations with surrogate NNs, unifying normal-operation "Resource Harvesting" with emergency "Urgent Response with Refinement."
Specifically, by developing \texttt{CPUGPU-Harvest} to leverage heterogeneous computing resources, we reduced energy-to-solution to 0.24 times that of conventional methods (a 76\% reduction) and improved throughput 3.7-fold. This realized effective resource harvesting that flexibly utilizes short idle backfill slots during daily supercomputer operations, enabling the construction of high-precision pretraining datasets at a practical cost of 263 node-hours. 
Next, by coupling \texttt{EBEGPU-Urgent}---which prioritizes time-to-solution during emergencies---with surrogate NNs, we cut computation time and resource consumption by over 97.9\% compared to conventional approaches.
This paves the way to deliver physically consistent, scientifically sound insights---specifically high-fidelity spatial time-history ground motion distributions reflecting strong 3D nonlinear responses in large urban areas---within emergency decision-making timelines of 12.1 minutes post-event (1st-order estimation) and within 25.3 minutes (refined estimation) while incorporating interactive refinement.
Future directions include further enhancing reliability by accounting for uncertainties in ground structures and more complex ground motion input conditions, as well as extending the framework to higher frequency ranges.
Furthermore, beyond directly contributing to rapid, scientifically grounded decision-making for disaster mitigation during large-scale earthquakes, this study establishes an HPC utilization model bridging operational efficiency ("Harvesting") and advanced decision-making ("Urgent Computing") in future exa-scale environments, offering broad applicability to real-time inverse and identification analysis across diverse nonlinear time-evolution problems.

\section*{Appendix: High‑fidelity physical model}

In the multi-spring model---one of the high-fidelity physical models used in 3D nonlinear ground amplification---incorporated in this study, physical properties are evaluated at four points per tetrahedral element using past history data to update the element stiffness matrix $\mathbf{K}_e$ composing $\mathbf{K}^n$ in Eqs.~\eqref{eq:GE} and \eqref{eq:newmarkbeta} as
$\mathbf{K}_e=\sum_{j=1}^5 w_j \mathbf{B}^T_{e,j} \mathbf{D}_{e,j} \mathbf{B}_{e,j}.$
Here, $\mathbf{B}^T_{e,j}$ is a $6 \times 30$ matrix converting nodal displacements into strains, $\mathbf{D}_{e,j}$ is a $6 \times 6$ elastoplastic stiffness matrix at the integration point, and $w_j$ represents the weight of the integration point.
In this study, this elastoplastic stiffness matrix $\mathbf{D}$ is calculated as 
$ \mathbf{D}=K \mathbf{m}\mathbf{m}^T + \sum_{i=1}^{N}w_i^s \frac{\text{d} \tau_i}{\text{d} \gamma_i} \mathbf{n}_i\mathbf{n}_i^T, $
where $K$ is the bulk modulus and $\mathbf{m}=\{1,1,1,0,0,0\}^T$.
Additionally, $\gamma_i$, $\tau_i$, and $w_i^s$ denote the strain, stress, and weight of the $i$-th 1D spring, respectively, and $\mathbf{n}_i$ is a vector converting 1D spring strain into 3D strain. In other words, the multi-spring model is a physical model that evaluates 3D time-history nonlinear behavior by combining numerous experimentally derived 1D nonlinear springs, where strain induces nonlinearity, leading to stiffness degradation and increased damping.
Here, because the modified Ramberg-Osgood model \cite{RO} and Masing's rule \cite{massing} are employed as the physical model for each 1D spring, 40 bytes of data—comprising four double-precision variables and two flags—must be maintained per 1D spring.
With the number of 1D springs set to $N=150$, a total of 24 KB of history data must be stored per tetrahedral element.
Thus, while this enables high-fidelity physical simulations, maintaining such large history data results in a memory and computationally expensive simulation.

\section*{Appendix: Neural network for estimating input earthquake motion}

While various methods such as PINNs and FNO exist, considering stable performance guaranteed learning from a small dataset based on domain knowledge of earthquake engineering, we constructed a CNN-LSTM encoder-decoder network that estimates input bedrock waveforms (3-component $x,y,z$) from time-history surface observation waveforms (3-component $x,y,z$) recorded at observation points.
As an architecture suited for handling both history dependence and dynamic locality/long-term temporal dependencies---widely used in waveform transformation tasks---we adopted a combination of CNN and LSTM \cite{CNNLSTM}.
Weight sharing in CNN enables efficient feature extraction in the time domain with fewer parameters, suppressing overfitting even when training data is limited \cite{CNN}.
First, target time-history waveforms were downsampled to the time domain corresponding to the target frequency range, yielding input and output time-history waveforms both of sequence length 1,813 with 3 components ($1813\times3$ dimensions).
The encoder transforms the input into latent features by stacking $n_c$ layers of 1D convolutional layers (kernel size $k$, stride 2) and ReLU activations, and extracts temporal dependencies using an LSTM ($n_{\text{LSTM}}$ layers, hidden dimension $n_{\text{hidden}}$, same as the size of latent features). 
The decoder features a symmetric transposed convolutional structure; however, non-linear activation is omitted in the final layer, mapping the 3 components via grouped convolution ($\text{groups}=3$), and the output length is matched to the input via linear interpolation.
As described in the main text, $NN^{P_2}$ was pre-constructed using 100 pairs of random wave inputs and their corresponding responses at observation point $P_2$.
Hyperparameters $n_c, n_{\text{LSTM}}, n_{\text{hidden}}, k$, and the learning rate were optimized using Optuna \cite{optuna} ($n_c\in\{2,3,4\}$, $n_{\text{LSTM}}\in\{1,2,3\}$, $n_{\text{hidden}}\in\{128,256,512,1024\}$, $k\in\{3,5,9,17,33,65\}$, learning rate $\in[5\times10^{-5},5\times10^{-4}]$, up to 200 epochs per trial) using $L_1$ loss and the Adam optimizer. 
Note that this was a single-objective optimization targeting the validation loss, conducted over 100 trials using the TPE sampler.
Using the 100 data pairs (80 training, 20 validation, $8:2$ random split), early stopping was applied if the validation loss did not improve for 20 epochs, adopting the best-performing model.
As a result of the optimization, $n_c = 2$, $n_{\mathrm{LSTM}} = 2$, $n_{\mathrm{hidden}} = 512$, $k = 3$, and learning rate $= 2.34 \times 10^{-4}$ were obtained. The relative L1 loss on the validation data decreased to $0.0206 \pm 0.0009$ at an average of 159.6 epochs (hereafter, values represent the average across 50 random seeds, reported alongside 95\% confidence intervals).
Implemented in PyTorch \cite{PyTorch}, training was conducted on 1 node of NVIDIA GH200---envisioning pre-construction across multiple domains using supercomputer backfill resources—performing mini---batch training with a batch size of 10, requiring 97.6 minutes including optimization.

To achieve event-specific adaptation, the pre-trained general-purpose model $NN^{P_2}$ is further trained on additional dataset ($s_{\text{sim}, P_2}^{*, (1)}$ and $b_{\text{est}}^{*, (1)}$) to construct $NN^{P_2}_{*, (1)}$.
To prevent catastrophic forgetting---where overfitting to additional data destroys the broad representation capability acquired by the surrogate NN during pre-training~\cite{overfit} ---$N_{\mathrm{rand}}=20$ random wave samples were mixed with $N_{\mathrm{aug}}=80$ duplicated copies of the additional training sample, trained using mini-batches of size 10 and early stopping with patience=20 (evaluated on the loss of additional training samples). 
Fine-tuning for the Kobe and Chuetsu-oki waves required an average of 1.21 minutes (82.0 epochs) and 1.64 minutes (115.5 epochs), respectively.
The achieved relative L1 errors were $0.0394 \pm 0.0021$ on the random wave validation data and $0.0328 \pm 0.0012$ on the additional training data for the Kobe wave, and $0.0307 \pm 0.0019$ on the random wave validation data and $0.0354 \pm 0.0014$ on the additional training data for the Chuetsu-oki wave.

\begin{figure}[tb]
\centering
\includegraphics[width=0.9\hsize]{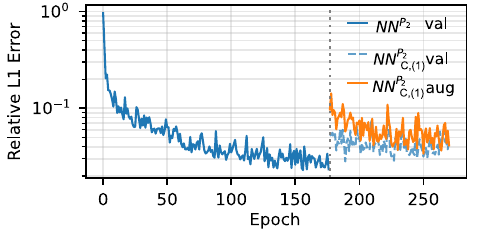}
\caption{\small{Combined training history of $NN^{P_2}$ and $NN^{P_2}_{\text{C}, (1)}$ (Chuetsu-oki scenario), separated by the dashed vertical line.}}
\label{fig:learning_hist}
\end{figure}

Figure~\ref{fig:learning_hist} shows the concatenated training history of $NN^{P_2}$ and fine-tuning history of $NN^{P_2}_{\text{C}, (1)}$, with a gray dashed line marking the boundary between the two (Chuetsu-oki scenario).
The blue line represents the error on the random wave validation data across both sides of the dashed line (left: $NN^{P_2}$, right: $NN^{P_2}_{\text{C}, (1)}$), while the orange line represents the error on the additional data for $NN^{P_2}_{\text{C}, (1)}$ on the right side of the dashed line.
While the fitting performance (orange line) improves through fine-tuning, the random wave validation error (blue line) degrades only slightly, confirming that generalization performance is not catastrophically impaired. It should be noted that the training/validation data consisted solely of random waves and $(s_{\text{sim}, P_2}^{\text{C}, (1)}, b_{\text{est}}^{\text{C}, (1)})$, and the test data (true values) was not used during training or hyperparameter tuning.

\section*{Acknowledgment}
The authors thank the Mainline Committee of the Association for the Development of Earthquake Prediction (ADEP) for providing the ground model used in this work. This work was supported by JSPS KAKENHI (Grant Numbers 26H02178, 25K21686).

\end{document}